\documentclass[a4paper,11pt]{article}
\pdfoutput=1 % if your are submitting a pdflatex (i.e. if you have
\usepackage{jinstpub} % for details on the use of the package, please
\usepackage{xcolor}
\usepackage{natbib}
\usepackage{graphicx}% Include figure files
\usepackage{dcolumn}% Align table columns on decimal point
\usepackage{bm}% bold math
\usepackage{hyperref}
\title{Characterization of Germanium Detectors for the First Underground Laboratory in Mexico.}% Force line breaks with \

\emailAdd{egarces@fisica.unam.mx}

\author[a]{A.~Aguilar-Arevalo}
\author[b]{S. Alvarado-Mijangos}
\author[c]{X. Bertou}
\author[d]{C. Canet}
\author[e]{M. A. Cruz-P\'erez}
\author[f]{A. Deisting}
\author[f]{A. Dias}
\author[a]{J. C. D'Olivo}
\author[a,c]{F. Favela-P\'erez}
\author[b,1]{E. A.Garc\'es\note{Corresponding Author.}}
\author[b,1]{A. Gonz\'alez Mu\~noz}
\author[a]{J. O. Guerra-Pulido}
\author[g]{J. Mancera-Alejandrez}
\author[b]{D. J. Mar\'in-L\'ambarri}
\author[a]{M. Martinez Montero}
\author[f]{J. Monroe}
\author[a]{C. Iv\'an Ortega-Hern\'andez} 
\author[h]{S. Paling}
\author[i]{S. Peeters}
\author[b]{D. Ru\'iz Esparza Rodr\'iguez}
\author[h]{P. R. Scovell}
\author[i,2]{C. T\"urko\u{g}lu\note{Now at :AstroCeNT - Particle Astrophysics Science And Technology Centre, ul. Rektorska 4, 00-614, Warsaw, Poland}}
\author[b]{E.~V\'azquez-J\'auregui}
\author[f]{J. Walding}

\affiliation[a]{Instituto de Ciencias Nucleares, Universidad Nacional Aut\'onoma de M\'exico, CDMX, M\'exico}
\affiliation[b]{Instituto de F\'isica, Universidad Nacional Aut\'onoma de M\'exico, A. P. 20-364, M\'exico D. F. 01000, Mexico}
\affiliation[c]{Centro At\'omico Bariloche,  CNEA/CONICET/IB, Bariloche, Argentina}
\affiliation[d]{Centro de Ciencias de la Atm\'osfera, Universidad Nacional Aut\'onoma de M\'exico, CDMX, 04110 Mexico}
\affiliation[e]{Programa de Posgrado en Ciencias de la Tierra, Universidad Nacional Aut\'onoma de M\'exico, Ciudad Universitaria, Coyoac\'an 04510, Ciudad de M\'exico, Mexico }
\affiliation[f]{Royal Holloway, University of London, Egham Hill, United Kingdom}
\affiliation[g]{Facultad de Ingenier\'ia, Universidad Nacional Aut\'onoma de M\'exico, Mexico}
\affiliation[h]{Boulby Underground Laboratory, Boulby Mine, Saltburn-by-the-Sea, United Kingdom}
\affiliation[i]{Department of Physics and Astronomy, University of Sussex, Brighton, United Kingdom}

\abstract{This article reports the characterization of two High Purity Germanium detectors performed by extracting and comparing their efficiencies using experimental data and Monte Carlo simulations. The efficiencies were calculated for pointlike $\gamma$-ray sources as well as for extended calibration sources. Characteristics of the detectors such as energy linearity, energy resolution and full energy peak efficiencies are reported from measurements performed on surface laboratories. The detectors will be deployed in a $\gamma$-ray assay facility that will be located in the first underground laboratory in Mexico, Laboratorio Subterr\'aneo de Mineral del Chico (LABChico), in the Comarca Minera UNESCO Global Geopark\,\cite{comihi}.\\
}
\begin{document}
\maketitle
\flushbottom

\section{\label{sec:intro}Introduction}

LABChico (Laboratorio Subterr\'aneo de Mineral del Chico) will be an underground laboratory in a decommissioned silver mine from colonial times, at the Comarca Minera UNESCO Global Geopark\,\cite{comihi}, in the Mexican state of Hidalgo. The LABChico research program will focus on applications for environmental radiation monitoring, $\gamma$-ray assay and screening, as well as prototype design for detectors in underground astroparticle physics experiments, see \,\cite{Lawson:2020ehz,Erchinger:2015ehf, Cleveland:2012sk,Ludwig:2019rdu,Szucs:2019awp, international1989iaea} for some examples.
This program requires a variety of state-of-the-art detector technologies which operate in a low radioactivity background environment, such as High Purity Germanium detectors. LABChico will be located at a depth of approximately 100\,m of rock, 300\,m.w.e. (meter water equivalent) overburden. The background cosmic ray flux is attenuated, from 1\,muon/cm$^2$/minute to approximately 1\,muon/cm$^2$/hour.

LABChico will consist of a 125\,m$^3$ cavern with a usable surface area of 37\,m$^2$ (including a platform), as well as a user/visitor center and storage facility outside the mine. The infrastructure will include ventilation and air conditioning (HVAC), electrical power, compressed air and networking. This work reports the characterization of two High Purity Germanium detectors that will be part of the LABChico underground facility.

This report is organized as follows. In section\,\ref{sec:GeDets} the detectors characterization with radioactive sources is presented; energy calibration and resolution are reported in sections\,\ref{sec:cal} and\,\ref{sec:res} respectively. The efficiency measurement is described in section\,\ref{sec:efficiency}. The simulation of the germanium detectors used in the efficiency measurement is described in section\,\ref{sec:simulation}. The validation with extended sources is described in section\,\ref{sec:ext}.

\section{\label{sec:GeDets} Germanium detector characterization}

The LABChico assay facility will initially consist of one High Purity Germanium detector manufactured by ORTEC, enclosed by a 5 inch thick lead shield. The detector has been characterized, for this work, at the Instituto de Ciencias Nucleares (ICN), Universidad Nacional Aut\'{o}noma de M\'exico (UNAM) in Mexico City. This detector will be hereafter referred to as ICN-HPGe, see figure\,\ref{Fig:ICNPhoto_tags}. The physics research and assay program will be complemented with a Broad Energy Germanium detector, located at the Instituto de F\'isica (IF), also at UNAM in Mexico City, manufactured by Canberra. This detector will be hereafter referred to as IF-BEGe, see figure\,\ref{Fig:IFPhoto_tags}. Both laboratories are on surface (Mexico City altitude 2,250 meters above mean sea level).

Both detectors are cooled down using liquid nitrogen, in a 30\,l cryogenic storage dewar. The ICN-HPGe has an operational bias of -3200\,V, with a p-type structure and a coaxial closed vertical geometry. The data acquisition system (DAQ) consists of a PX5-HPGe multichannel analyzer (MCA) and the digital pulse processor (DPP) software analyzer provided by Amptek\,\cite{amptek}. 

The dimensions of the internal components of the ICN-HPGe detector were estimated from X-ray images and by measuring the external parts of the device (see figure\,\ref{Fig:ICNPhoto_tags}). The  identified components included in the simulations (see figure \ref{fig:HPGeDimensions}, section \ref{sec:simulation}) are:

\begin{figure}
     \centering
     \includegraphics[width=0.3\textwidth]{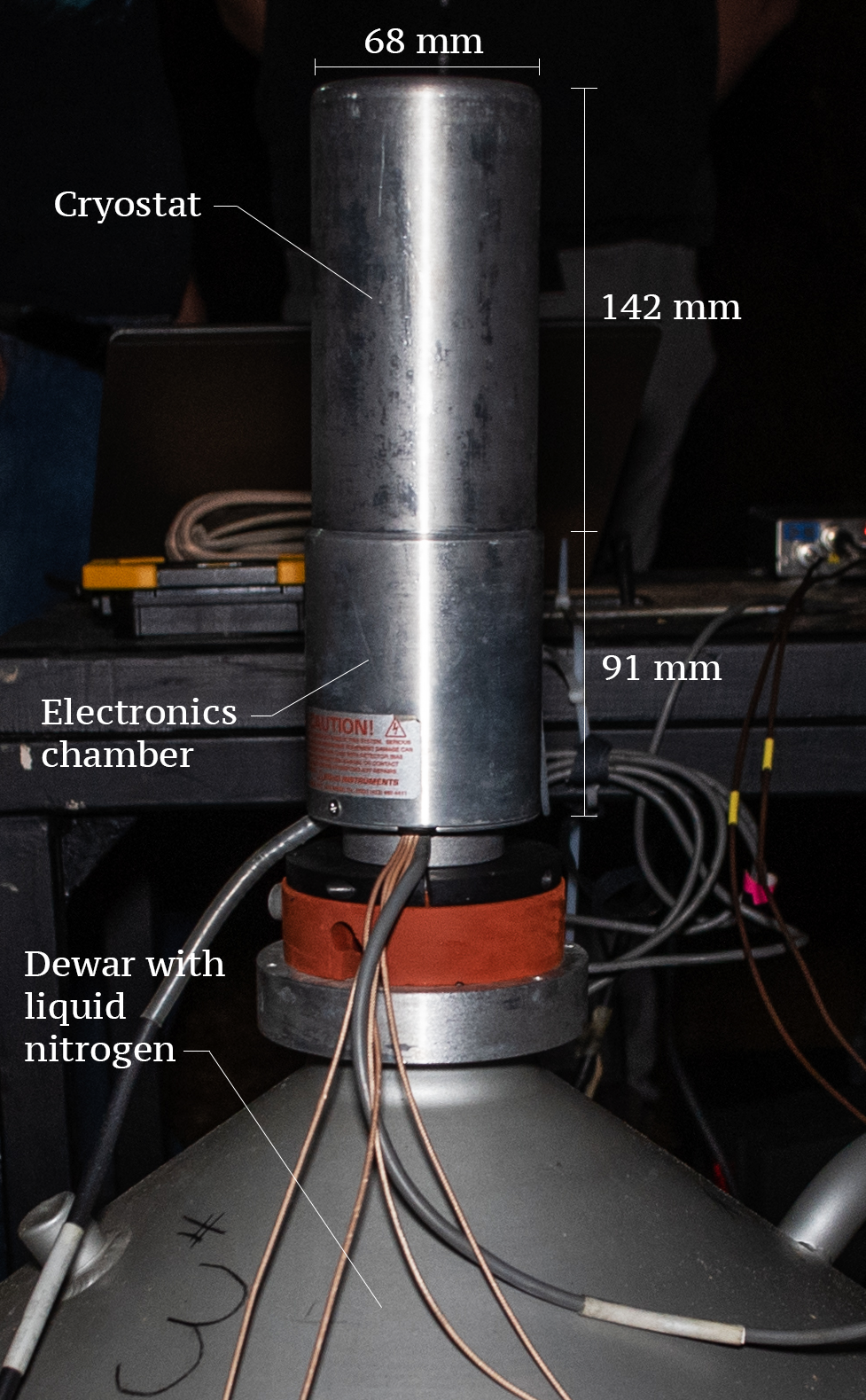}
     \caption{The High Purity Germanium detector (ICN-HPGe) with its liquid nitrogen dewar at the Instituto de Ciencias Nucleares, UNAM.}
     \label{Fig:ICNPhoto_tags}
 \end{figure}

\begin{enumerate}
    \item Cryostat: an aluminium cylinder of 34.0\,mm radius, 142.0\,mm height and 500\,$\mu$m thickness. It contains the germanium crystal, contacts and endcap.
    \item Outer contact: lithium cylinder of 700\,$\mu$m thickness, surrounding the germanium crystal.
    \item Dead layer: this is the part of the germanium crystal that is not sensitive to the incoming $\gamma$-rays. Its thickness was determined using Monte Carlo simulations as described in section \ref{sec:simulation}.
    \item Germanium crystal: a cylinder of 25.5\,mm radius and 54.0\,mm height with a cylindrical hole in the middle for the inner contact.
    \item Endcap: made of carbon fiber with a 1.0\,mm width aluminium window to allow low energy photons to interact with the sensitive crystal.
    \item Thermal strap: an aluminum disc of 120\,mm radius.
    \item Electronics chamber: an aluminum cylinder of 39.0\,mm radius, 70.0\,mm height and 500\,$\mu$m width that contains the preamplifier electronics.
    \item Inner contact: a 0.3\,$\mu$m inner contact of boron implanted ions.
    \item Cold finger: an aluminum cylinder of 7.50\,mm radius.
\end{enumerate}

The IF-BEGe is a Canberra model BE2820\,\cite{Canberra} with planar horizontal crystal configuration, 30\,mm radius and 20\,mm height, an operational bias of +3000\,V, a DAQ consisting of a Pocket MCA 8000\,A and the ADMCA software also provided by Amptek\,\cite{amptek}, see figure\,\ref{Fig:IFPhoto_tags}. 

For the IF-BEGe detector the dimensions were obtained from the technical sheet provided by the manufacturer and also measuring the external parts of the device (see figure\,\ref{Fig:IFPhoto_tags}). The components identified and later included in the simulations (see figure\,\ref{fig:geometry}, section\,\ref{sec:simulation}) are:

\begin{figure}
     \centering
     \includegraphics[width=0.5\textwidth]{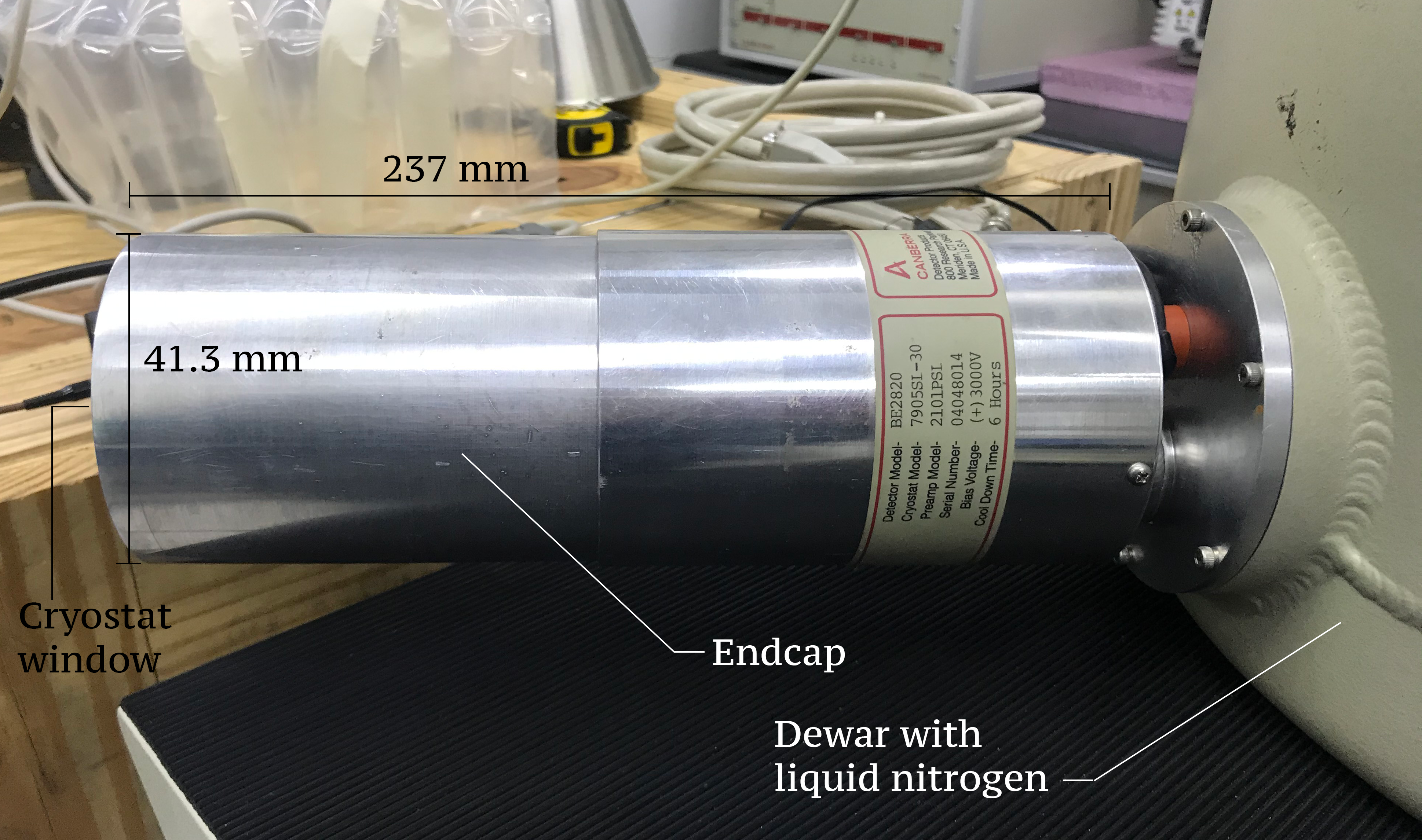}
     \caption{The Broad Energy Germanium detector (IF-BEGe) at the Instituto de F\'isica, UNAM.}
     \label{Fig:IFPhoto_tags}
 \end{figure}

\begin{enumerate}
  \item Cryostat window: a disk with a 32.0\,mm radius and 0.6\,mm thickness, made of carbon composite.
  \item Infrared (IR) window: two thin disks of 30.0\,mm radius. One made of polyethylene and 7.4\,$\mu$m thickness and the other made of aluminum and 0.1\,$\mu$m thickness.
 \item Side electrode dead layer: region of the germanium crystal, in contact with the electrode, unable to detect $\gamma$-rays.
 \item Germanium crystal: a germanium cylinder of 28.55\,mm of radius and 17.15\,mm height.
  \item Teflon cup: this volume is located between the detector holder and the germanium crystal. It is a cup with inner radius 30.1\,mm, outer radius 30.6\,mm and  25.5\,mm height.
 \item Vacuum space: inside the Cryostat a vacuum environment protects the germanium crystal surfaces from moisture and condensible contaminants.	 
  \item Endcap: an aluminum cylinder with  41.3\,mm radius, 237.0\,mm long and 1.6\,mm thickness.
 \item Front electrode dead layer: region of the germanium crystal unable to detect $\gamma$-rays.
  \item Detector holder: This volume is made of copper and has the shape of a cup, with inner radius of 30.6\,mm and outer radius of 31.4\,mm. In the top side of this cup there is a ring that widens the outer radius up to 33.6\,mm. The height of the holder is 27.1\,mm.
 % \item \textcolor{blue}{Front electrode dead layer: }
%  \item \textcolor{blue}{Detector holder: }
\end{enumerate}

The main difference between the ICN-HPGe and the IF-BEGe detectors is the geometry of the germanium crystals. The ICN-HPGe has a bulletized coaxial crystal, that is, a crystal in which the corners facing the front of the detector have been rounded to avoid charge collection in regions where the electric field is weak\,\cite{GGilmoure}. The IF-BEGe detector has a planar crystal in which the applied electric field is more uniform than in a coaxial crystal. These differences have an effect in the charge collection produced by $\gamma$-ray interactions which become visible in the detector response.

 \subsection{Energy linearity \label{sec:cal}}
 
A set of pointlike radioactive sources, provided by \textit{Spectrum Techniques}\,\cite{LLC}, described in table\,\ref{Tab:sources}, have been used to calibrate the energy response of the detectors. A 20\,\% uncertainty on the source activities is reported by the manufacturer.

 \begin{table}[h]
\centering
\begin{tabular}{c|c|c}
\hline
 Source & $\gamma$ peaks [keV]  & Half-life [yr]\\
 \hline \hline
 $^{241}$Am & 59.54  & 432.2  \\ \hline
 $^{133}$Ba & 81.0, 276.0, 303.0, 356.0 & 10.5 \\ \hline
  $^{109}$Cd & 88 & 1.27 \\ \hline
 $^{57}$Co & 122.0, 136.0 & 0.745 \\ \hline
 $^{22}$Na & 511.0, 1275.0 & 2.6 \\ \hline
 $^{137}$Cs & 662.0 & 30.1 \\
 \hline
 $^{54}$Mn & 835.0 & 0.855 \\
 \hline
 $^{65}$Zn & 1115.0 & 0.668 \\
 \hline
 $^{60}$Co & 1173.2, 1332.5 & 5.27 \\
 \hline
\end{tabular}
\caption{\label{Tab:sources} Energy and half-life of the $\gamma$-ray sources. All having an initial activity (January 2019) of 1\,$\mu$Ci and a 20\,\% uncertainty, except for $^{137}$Cs (0.1\,$\mu$Ci)\,\cite{LLC}.}
\end{table}

\begin{figure}
    \centering
    \includegraphics[scale=0.36]{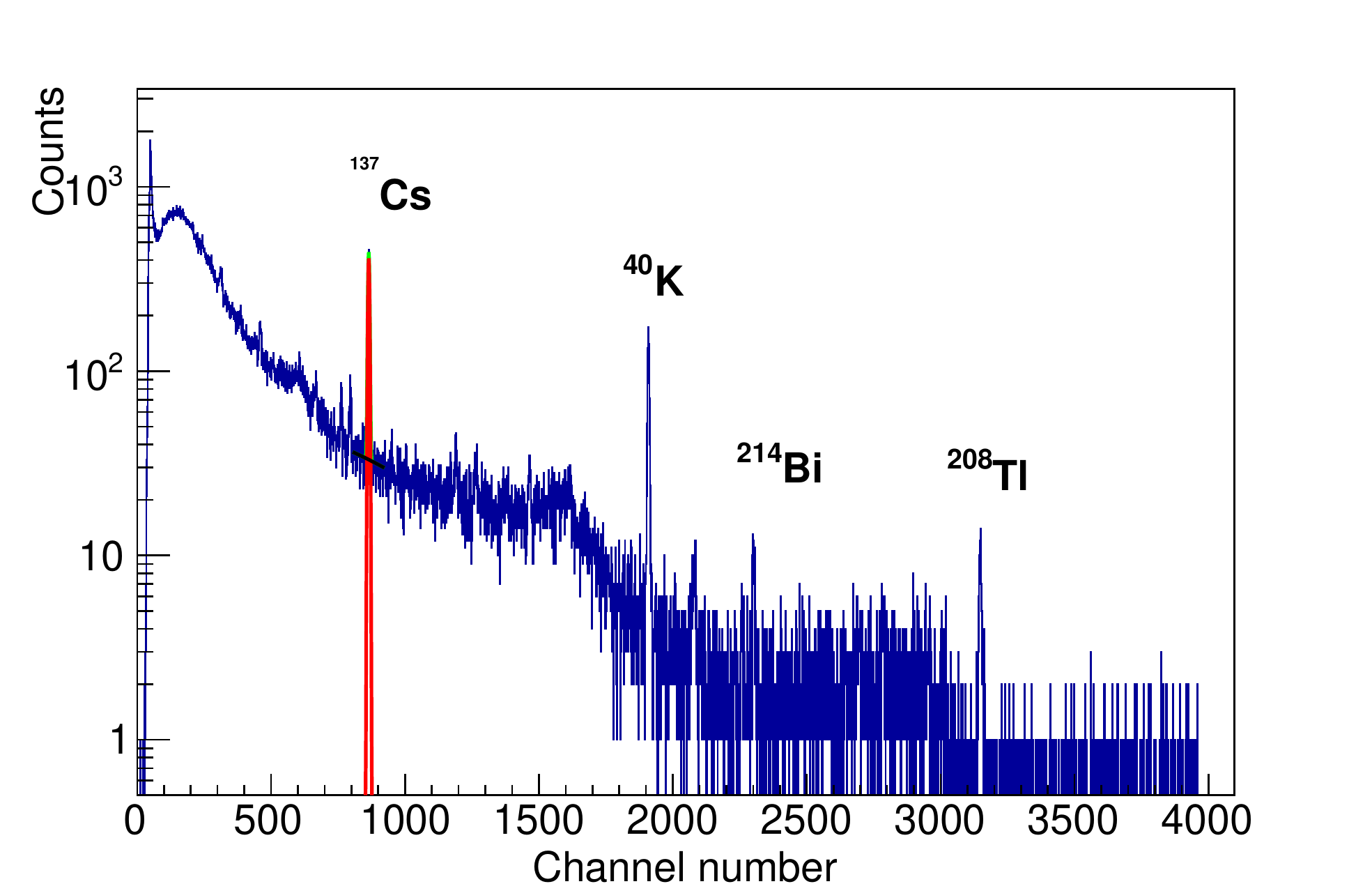}      \includegraphics[scale=0.36]{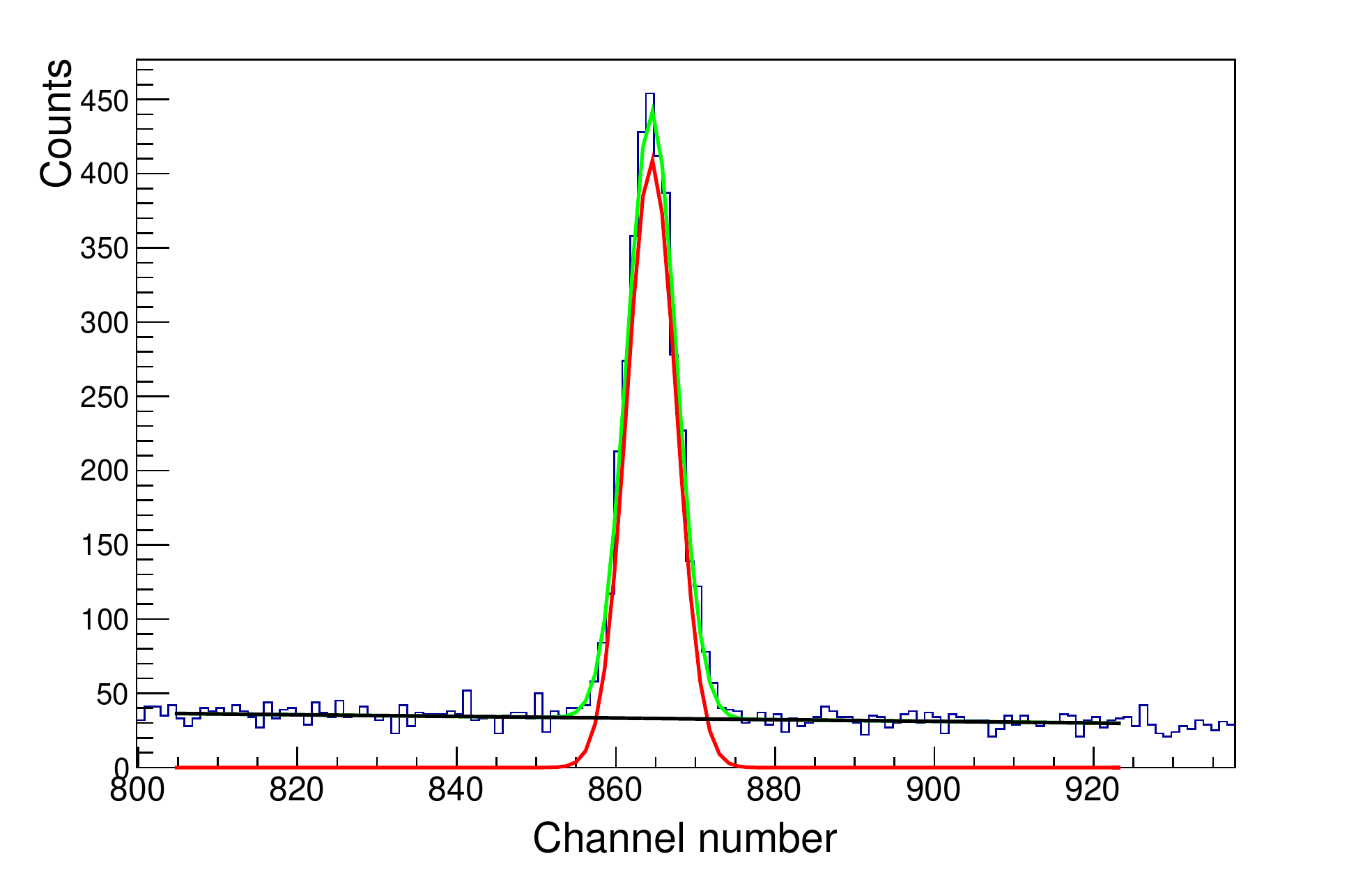}
    \caption{Raw data spectrum of the IF-BEGe detector placing the $^{137}$Cs radioactive source in front of the detector, in the vertical axis are shown the number of counts per channel and in the horizontal axis is the channel number to which an energy is associated. Left panel shows the full spectrum range, 4096 channels equivalent to up to 3\,MeV,the $^{137}$Cs photopeak is shown in red. The other peaks observed in the energy spectrum, $^{214}$Bi and $^{208}$Tl, are decay products of the primordial radionuclides $^{238}$U and $^{232}$Th, respectively. $^{40}$K is also a primordial radionuclide. These naturally occurring isotopes are ubiquitously present in nature. Right panel is a zoom around the peak corresponding to the $^{137}$Cs photopeak, the red line is the fitted Gaussian function, the black line is the order one polynomial and the green line shows the fitted function composed by the sum of both, the fit is performed in a six sigma range about the photopeak mean.}
    \label{Fig:spec}
\end{figure}

\begin{figure}
    \centering
    \includegraphics[scale=0.4]{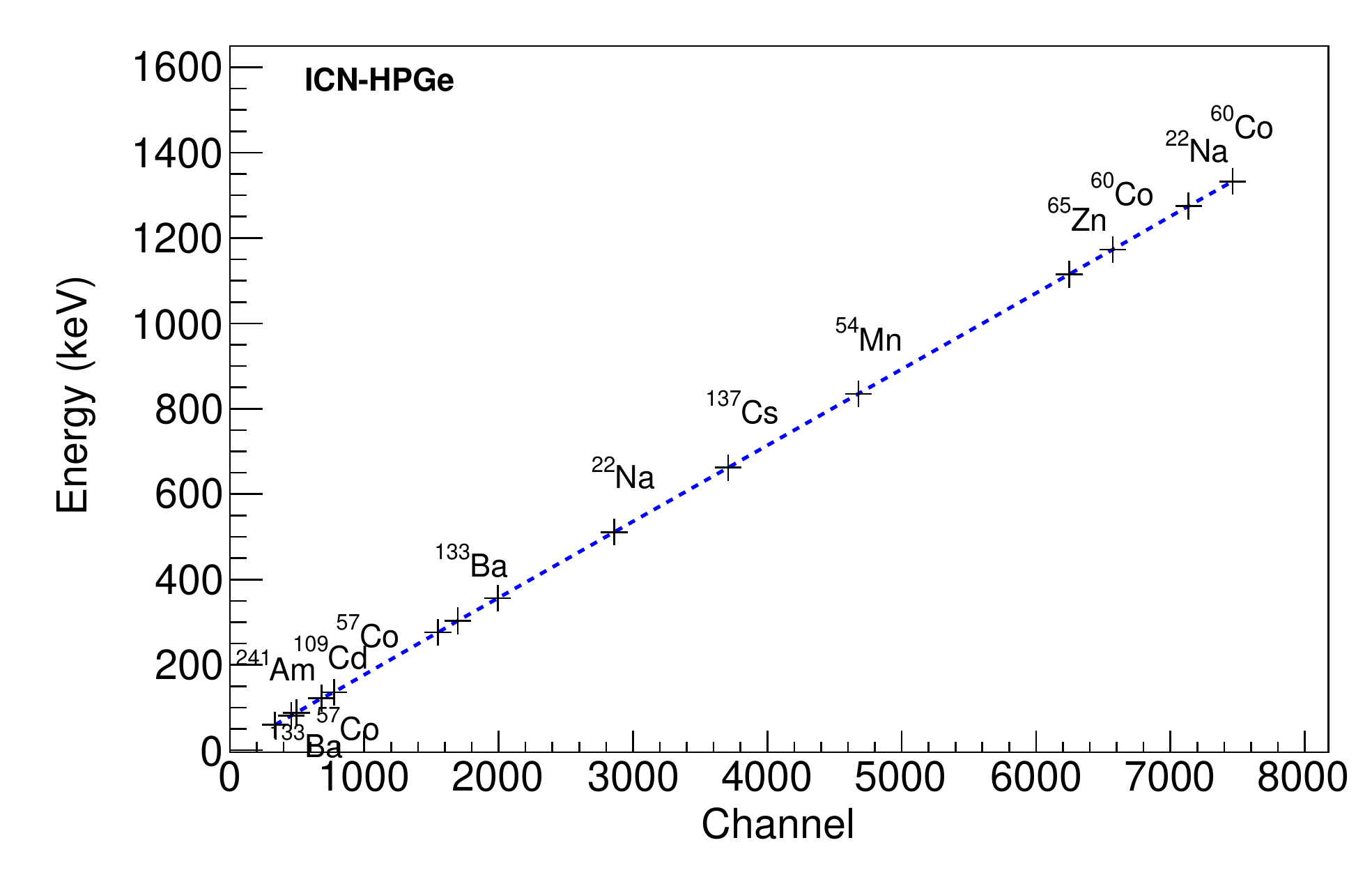}
      \includegraphics[scale=0.4]{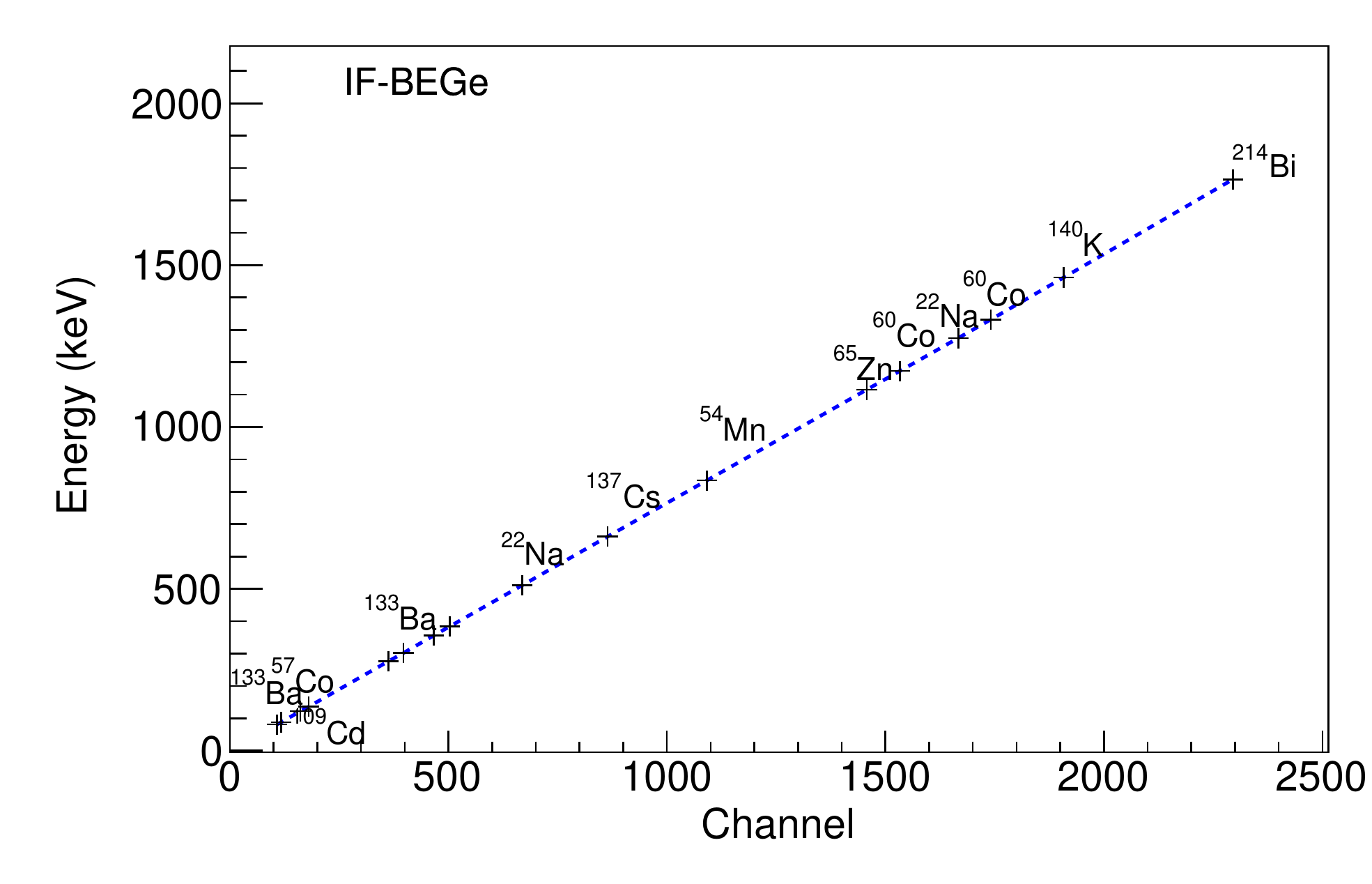}
    \caption{Linear energy response of the detectors for different $\gamma$-ray sources. Deposited energy (keV) vs. channel number. In the top panel is presented the ICN-HPGe detector calibration, in the bottom panel the calibration curve for the IF-BeGe detector. Fitting parameters of the linear functions can be found in table \ref{Tab:lin}.}
    \label{Fig:cal}
\end{figure}

Each source was deployed individually, at a distance of 25\,cm from the detector endcap along its symmetry axis. The measurements with the ICN-HPGe detector were performed with the source and the detector enclosed by the lead shield, whereas the measurements with the IF-BEGe detector were performed without shield. A spectrum was acquired for each source; an example of the $^{137}$Cs spectrum is shown in figure\,\ref{Fig:spec}, left panel. For each spectrum, the photopeak was fitted to a Gaussian plus a polynomial order one function using ROOT analysis tools\,\cite{BRUN199781}. The Gaussian mean fit parameter is the channel value associated to the $\gamma$-ray energy and its sigma is related to the detector energy resolution presented in section\,\ref{sec:res}, see figure\,\ref{Fig:spec}, right panel. The integral of the fitted linear function background model is subtracted from the total number of counts in a six sigma range about the mean. From this exercise with each spectrum the linear distribution of $\gamma$-ray energy vs. channel is obtained. The fit parameters of the energy response distributions shown in figure\,\ref{Fig:cal}, are listed in table \ref{Tab:lin}\footnote{For the IF-BEGe detector, if the b parameter is fixed to zero there is no significant change in the slope parameter m.}.

For the IF-BEGe detector, an extra two natural background radiation points ($^{40}$K 1.46\,MeV  and $^{214}$Bi 1.76\,MeV) are present in the spectrum and were included for the energy calibration, also $^{208}$Tl peak is present at 2.6\,MeV; these can be seen in the left panel of figure~\ref{Fig:spec}. 

\begin{table}[h]\centering
\begin{tabular}{c|c|c}
\hline
 Parameter & ICN-HPGe & IF-BEGe \\
 \hline \hline
 m &   0.1800 $\pm$ 0.0001 & 0.7662 $\pm$ 0.0002 \\
b &  -0.19 $\pm$ 0.32  & -1.62 $\pm$ 0.19 \\
\hline 
\end{tabular}
\caption{\label{Tab:lin} Fit parameters for the detector calibrations in figure \ref{Fig:cal}, energy $E$ as a function of the associated channel $N$ is $E(N) = m\times N +b$.}
\end{table}

\subsection{Energy Resolution \label{sec:res}}

\begin{figure}[!t]
    \centering
    \includegraphics[scale=0.4]{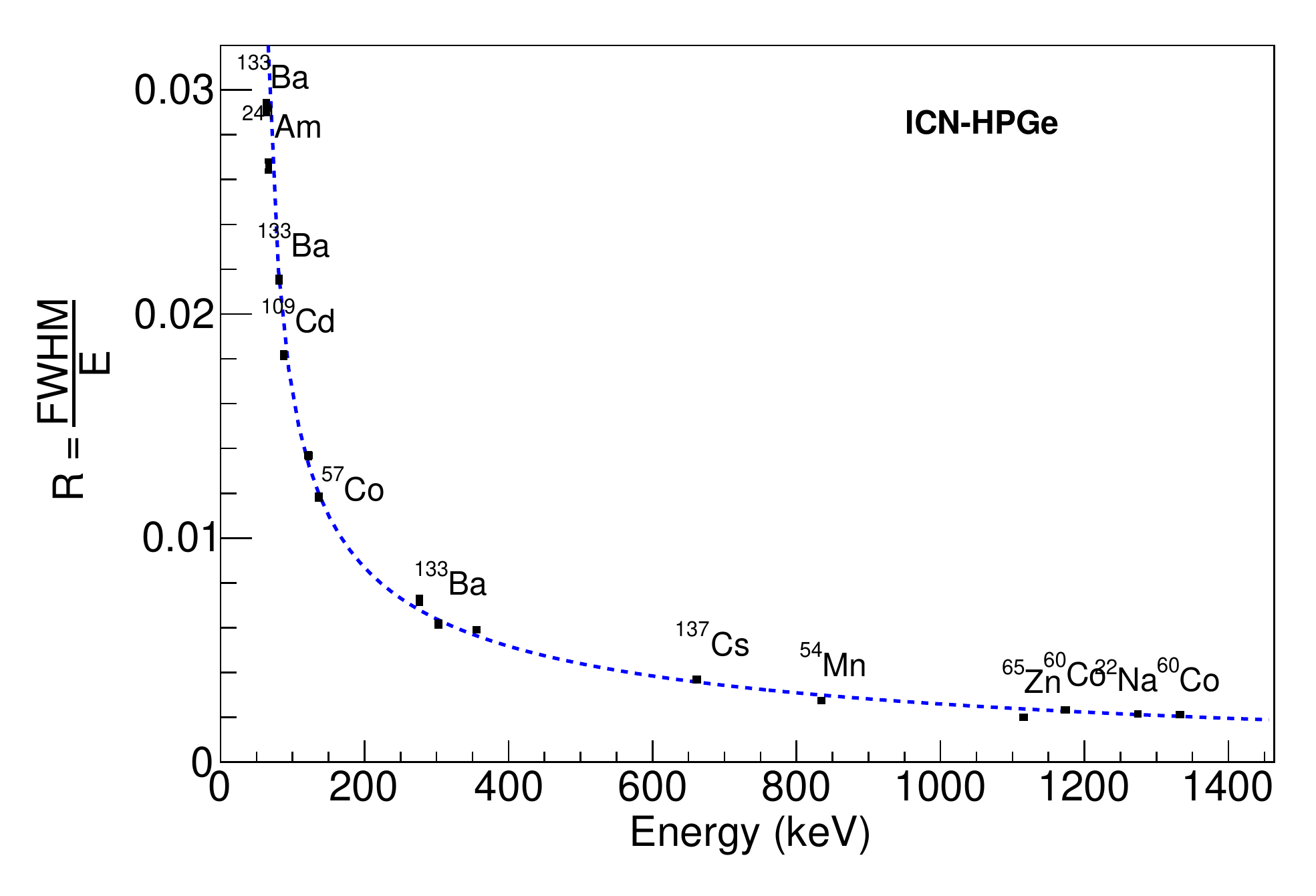}
      \includegraphics[scale=0.4]{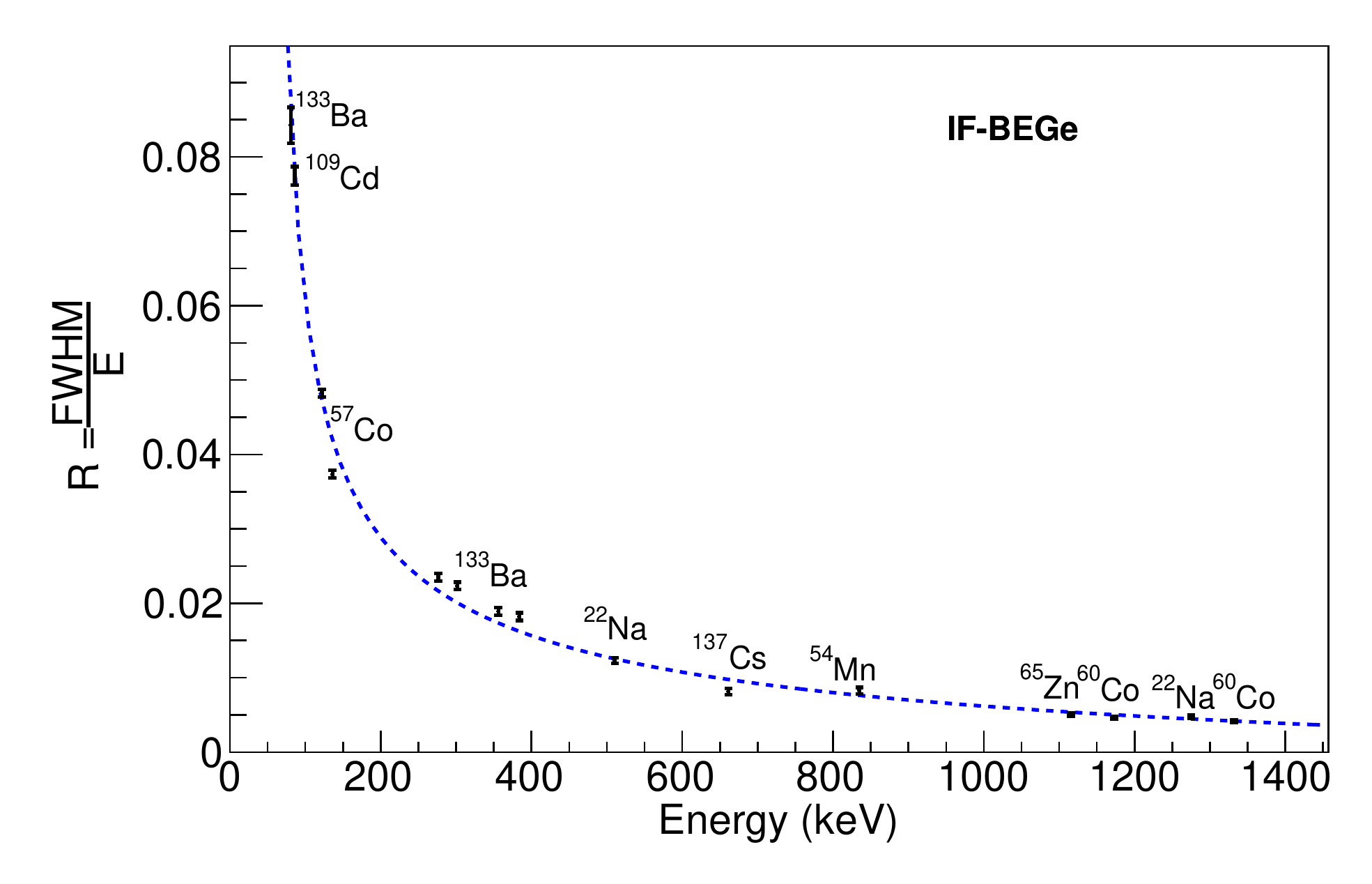}
    \caption{\label{Fig:Res} Resolution as a function of gamma energy, $R(E)=[P_0]/\sqrt{[P_1]+E} + [P_2]$. The top panel shows the resolution curve vs. energy for the ICN-HPGe detector, and bottom panel the resolution curve for the IF-BEGe detector. The ICN-HPGe shows a better resolution performance compared to that of the IF-BEGe, which does not meet the expected resolution performance from factory settings\,\cite{Canberra}, having almost twice the expected FWHM. The ICN-HPGe shows an energy resolution at 1.3 MeV compatible with values reported by ORTEC\,\cite{ortec_res} for similar detectors. Fit parameters can be found in table\,\ref{Tab:res} for both detectors. }
\end{figure}

After the energy scale calibration, the spectrum is fitted to a Gaussian function plus an order one polynomial with negative slope. The full width at half maximum (FWHM) is given as $FWHM=2\sqrt{2\ln{2}}\times\sigma \simeq 2.355\sigma$, where  $\sigma$[keV] is the Gaussian fit parameter to each photopeak. The detector energy resolution is defined as the ratio of the  FWHM to the true gamma peak energy, $R = {FWHM}/{E}$, the uncertainty in $R$ is obtained from the uncertainty in $\sigma$ and E, which are obtained from the fit\footnote{The errors in $\sigma$ and E were augmented by a factor $\chi^2$/n.d.f. given by the fit to account for non-Gaussianities in the measured photopeaks when this number was greater than one, for $\chi^2$/n.d.f. smaller than one, this correction was not applied.}. This distribution is shown in figure\,\ref{Fig:Res} for both detectors, fitted to an empirical three parameter inverse square root function\,\cite{grupen}: 
\begin{equation}\label{Eq:ResE}
    R = \frac{[P_0]}{\sqrt{[P_1]+E}}+[P_2],
\end{equation} 
the best-fit values to the fitted parameters $P_0$, $P_1$ and $P_2$, in [keV], are listed in table~\ref{Tab:res}.

\begin{table}[h]\centering
\begin{tabular}{c|c|c}
\hline
 Parameter & ICN-HPGe & IF-BEGe\\ \hline
 \hline 
$P_0$ (keV)& 1.38$\times$10$^{-1}\pm$1.14$\times$10$^{-2}$& 4.49 $\times$10$^{-1}\pm$1.0$\times$10$^{-3}$\\
 \hline
$P_1$ (keV)& -36.73$\pm$5.21  & -57.68$\pm$0.11 \\
 \hline
$P_2$ & -1.63$\times10^{-3}\pm$3.54$\times10^{-4}$  & -8.45$\times10^{-3}\pm$2.70$\times10^{-5}$ \\
\hline 
\end{tabular}
\caption{\label{Tab:res} Fit parameters for the detectors resolution function, Equation \ref{Eq:ResE}.}
\end{table}

\subsection{\label{sec:efficiency}Efficiency}

Absolute efficiency, also known as full energy peak efficiency, is defined as the ratio of the number of counts detected in a peak to the total number emitted by the source, 
\begin{equation}
\epsilon=\frac{N_{FEP}}{P_{\gamma} N_{TOT}},
\label{eq:eff}
\end{equation}
where $N_{FEP}$ is the  full energy peak count rate in counts per second, $P_{\gamma}$\footnote{ {$P_\gamma$ is the fraction of $\gamma$ emissions of a given energy to the total number of isotope disintegrations, this values and their corresponding uncertainties are taken from\,\cite{NLHB}.}} is the emission probability of the $\gamma$-ray being measured, and $N_{TOT}$ is the total number of $\gamma$-rays emitted at the specific energy, which was corrected for decay from the date of preparation, see table~\ref{Tab:sources}. 

The total number of counts in each full energy peak has been computed by integrating the fitted function in a standard interval of six times the standard deviation (see right panel of figure\,\ref{Fig:spec}), which is symmetric about the mean of each photopeak. Counts below the straight line, used to fit the background, have been subtracted. 

\begin{figure}[t]
     \centering
     \includegraphics[scale=0.44]{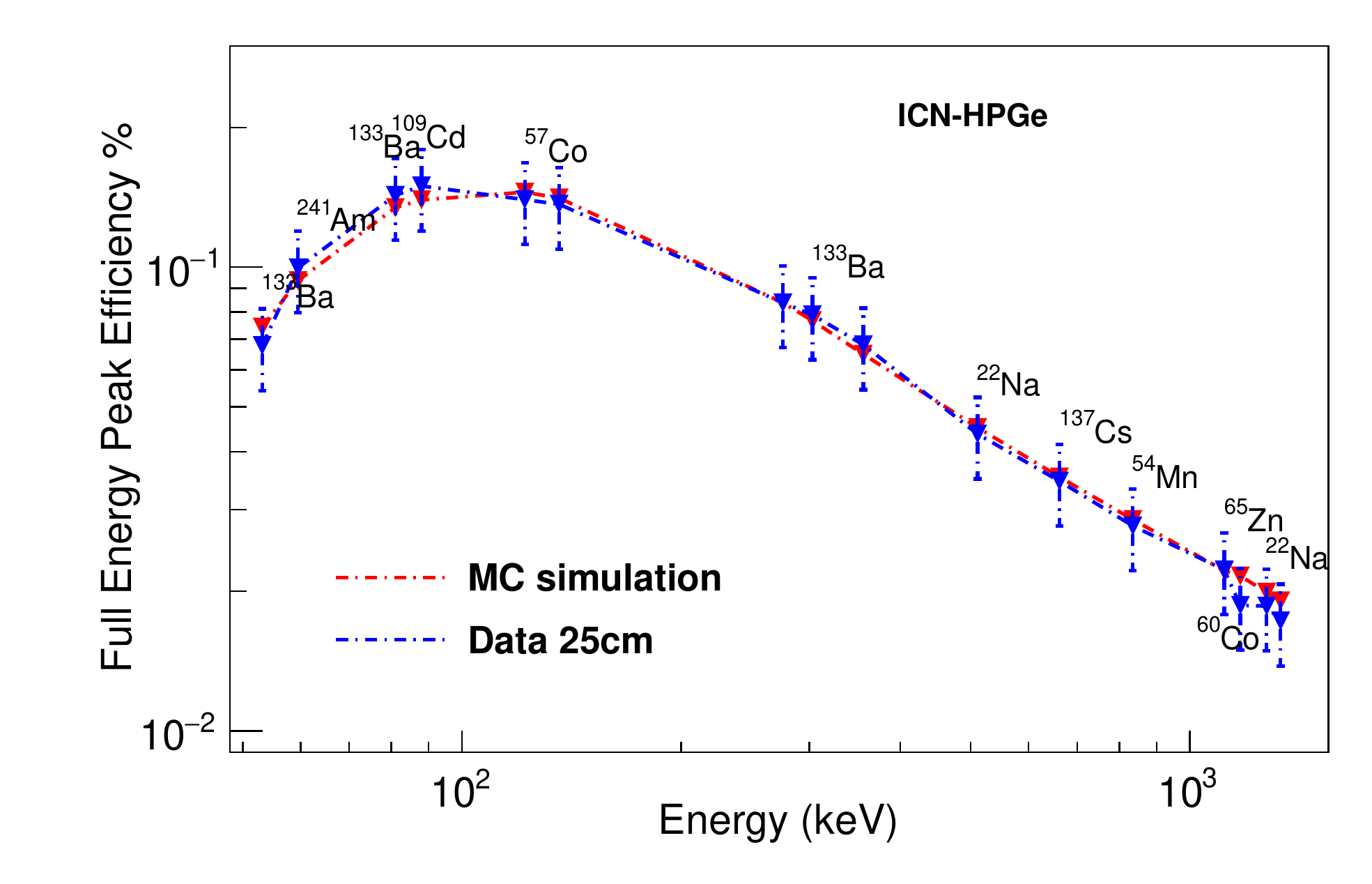}
      \includegraphics[scale=0.44]{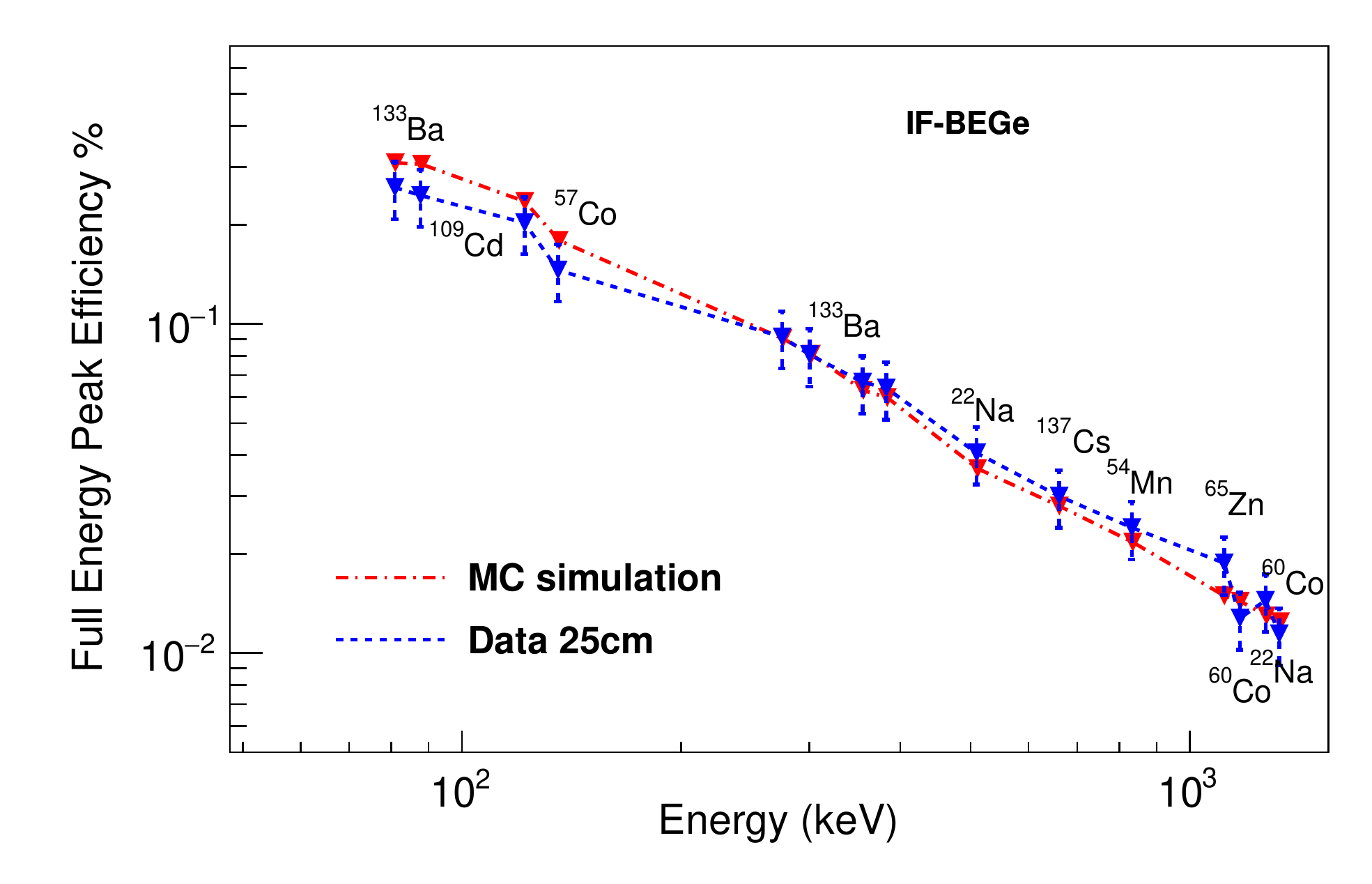}
     \caption{Germanium detectors efficiencies measured with pointlike $\gamma$-ray sources, 25\,cm away from the endcap of the detectors, compared with Monte Carlo simulations. Top panel shows the efficiency curve for the ICN-HPGe detector and bottom panel shows the IF-BEGe detector efficiency curve. }
     \label{Fig:HPGeEff}
 \end{figure}

The number of emitted $\gamma$-rays for each source is determined using the radioactive decay law and the age of the source. The half-life source values are presented in table\,\ref{Tab:sources} and the branching ratios for each $\gamma$-ray were taken from the National Laboratory Henri Becquerel decay tables\,\cite{NLHB}.

Figure \ref{Fig:HPGeEff} shows the efficiency measurements for all photopeak energies at a distance of 25\,cm. For the ICN-HPGe detector the expected behavior was observed: a fast rise of the efficiency from low energies up to a maximum expected around 100\,keV, between the $^{109}$Cd and $^{57}$Co $\gamma$-rays, and then a slower steady decrease towards high energies, which is consistent with similar detectors by ORTEC\,\cite{ortec_tech}\,\footnote{Due to a lack of documentation for the ICN-HPGe detector, a comparison of relative efficiency values with manufacturer specifications is not possible}. Uncertainties in figure\,\ref{Fig:HPGeEff} are assessed assuming uncorrelated errors from branching ratios, half-lives of the isotopes\,\cite{NLHB}, number of counts in the photopeak (statistical) and source activities (20\,\%).

Broad Energy Germanium detectors have a typical relative efficiency that can range from 9\,\% to up to 50\,\% depending on crystal volume and front face area\,\cite{Canberra}.  
For the IF-BEGe detector, model BE2820, a 13\,\% relative efficiency at 1332\,keV from $^{60}$Co is reported by Canberra\,\cite{Canberra}. For historical reasons, relative detection efficiency of germanium detectors is defined at 1.33\,MeV relative to the absolute efficiency of a standard NaI(Tl) scintillator, this standard is a crystal of 3\,in diameter and 3\,in long using a $^{60}$Co source placed 25\,cm from the endcap face which value is 1.2$\times$10$^{-3}$\,\cite{GGilmoure}.  The germanium detector full energy peak efficiency measured in this conditions divided by 1.2$\times$10$^{-3}$ is the relative efficiency specification of germanium detectors. The 13\,\% relative efficiency value reported for the IF-BEGe detector is equivalent to a full energy peak efficiency of 0.0156\,\%, and the measured value, (0.0117 $\pm$ 0.0033)\,\%, differs from the Canberra reported value by 25\,\%.

\section{\label{sec:simulation}Monte Carlo simulations}

Monte Carlo simulations for both detectors were performed using the simulation toolkit GEANT4\,\cite{AGOSTINELLI2003250}, 10.01.p03 version. The simulations included all the geometric elements that affect the propagation of $\gamma$-rays between the source and the germanium crystal.

\subsection{\label{sec:sim-HPGe}ICN-HPGe}
\begin{figure}[!t]
    \centering
    \includegraphics[width=0.5\textwidth]{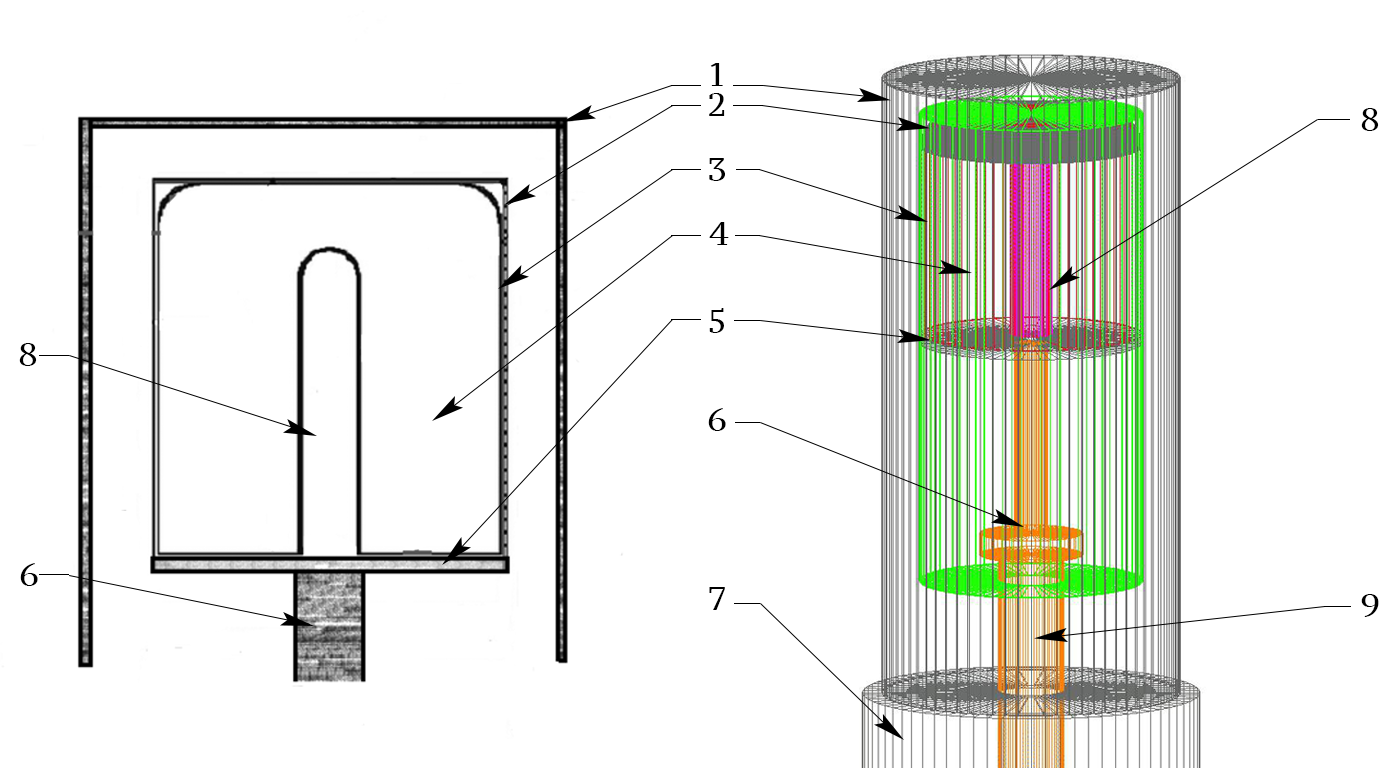}
    \caption{Left panel, the layout of the ICN-HPGe detector taken from a X-ray scan and right panel, its implementation in GEANT4. The components are: 1) Cryostat, 2) Outer contact, 3) Dead layer, 4) Germanium crystal, 5) Carbon fiber endcap, 6) Thermal strap, 7) Electronics chamber, 8) Inner contact and 9) Cold finger.}
    \label{fig:HPGeDimensions}
\end{figure}

For the ICN-HPGe detector, the components listed in section \ref{sec:GeDets} were included in the simulations (see figure \,\ref{fig:HPGeDimensions}). The preamplifier electronics (inside the electronics chamber) was not simulated.

There was no information available about the dead layer thickness surrounding the active volume of the germanium crystal. The X-ray images did not provide information about this parameter either. Dead layer thicknesses of the order of 0.75\,mm have been measured for a similar coaxial vertical germanium detectors manufactured by ORTEC by bombarding its crystal with a collimated $\gamma$-ray source\,\cite{ColSource}. Having similar dimensions, the ICN-HPGe detector crystal was expected to have a dead layer of around the same order. Monte Carlo simulations with varying dead layer thicknesses were performed to estimate this parameter by comparing the measured and simulated pointlike source photopeak efficiencies, a best match was found at a dead layer thickness of 0.65\,mm.

\subsection{\label{sec:sim-BEGe}IF-BEGe}
\begin{figure}
    \centering
    \includegraphics[width=0.5\textwidth]{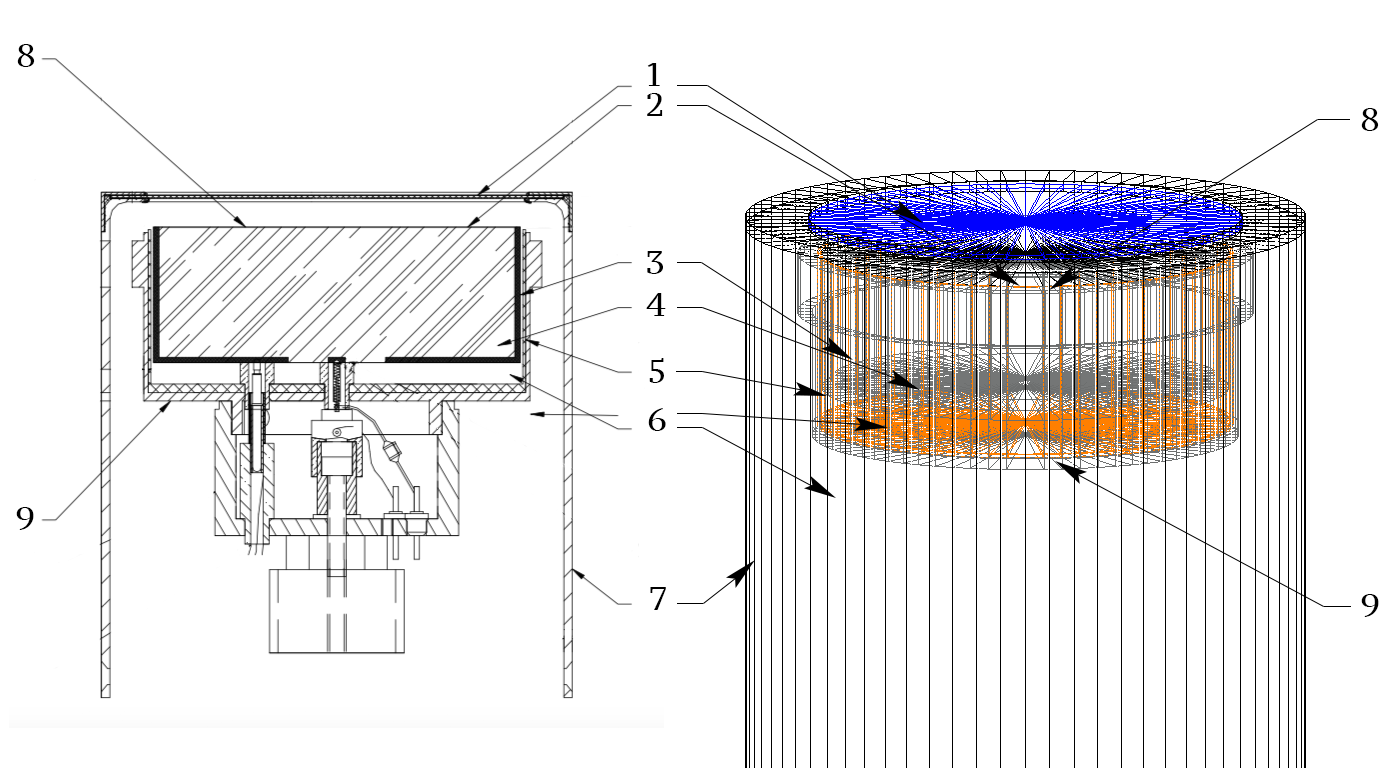}
    \caption{Geometry of the IF-BEGe detector. Left side, the detector cross-section provided by the manufacturer (Canberra)\,\cite{Canberra} and right side, the GEANT4 simulation of the detector. The components included in the simulations are: 1) Cryostat window, 2) IR window, 3) Side electrode dead layer, 4) Germanium crystal, 5) Teflon cup, 6) Vacuum space, 7) Endcap, 8) Front electrode dead layer and 9) Detector holder.}\label{fig:geometry}
\end{figure}

The IF-BEGe detector was simulated including the components and dimensions listed in section\,\ref{sec:GeDets} (see figure\,\ref{fig:geometry}). The dead layer thickness of the germanium crystal was modeled as 0.05\,mm on the front, 1.45\,mm on the sides and 2.8\,mm on the back, after the study described in the following section.

%\subsubsection{\label{sec:crystalScan}Crystal active volume scan}
A dead layer on the front side of the germanium crystal of  0.3\,$\mu$m and 500\,$\mu$m on the sides is reported by Canberra in the technical sheet of the IF-BEGe detector.  These settings were used in the simulations as a first approach, pointlike $\gamma$-ray sources located 25\,cm from the detector cryostat window were simulated. The efficiencies for these simulated sources were calculated similarly to those in section\,\ref{sec:efficiency}. A disagreement greater than 30\,\% was found when comparing the efficiencies obtained in the simulation with the experimental ones. Increasing the dead layer on the front side reported by the Canberra technical sheet up to two orders of magnitude was not sufficient to find an agreement between experimental data and the Monte Carlo simulations. 

The discrepancy between data and simulated efficiencies could be explained if the crystal active volume had a major alteration with respect to that reported in the technical sheet. In order to explore this possibility, a crystal scan was performed following a similar method as in\,\cite{Zeng:2016mef,Hult:2019nvv}.

\begin{figure}[t] \centering
     \includegraphics[scale=0.30]{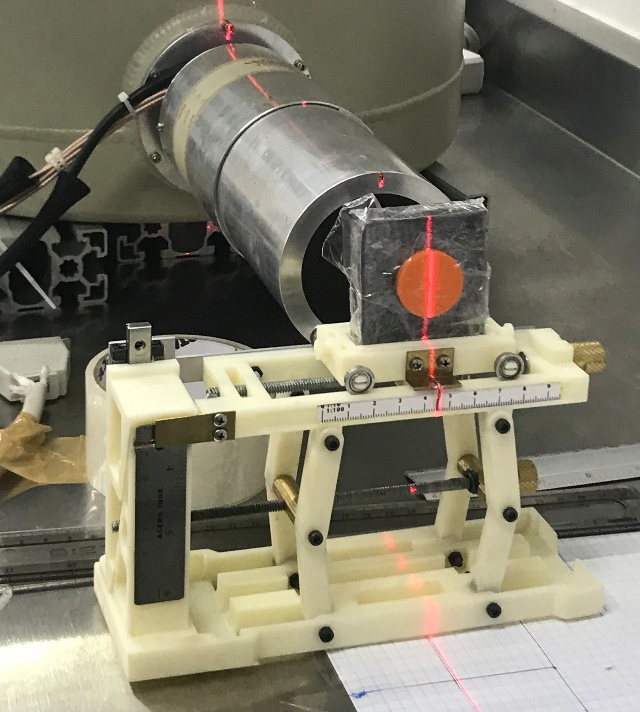}
     \caption{\label{Fig:detscan} Experimental setup for the IF-BEGe detector scan. The mechanical translation stage, $\gamma$-ray source (orange disk), lead block (collimator) and detector endcap are shown.}
 \end{figure}

For the scan, a pointlike $\gamma$-ray source ($^{109}$Cd) was fixed on top of a mechanical translation stage, in order to move it in a plane parallel and 4~cm away from the frontal face of the germanium crystal. The length of the mechanical translation stage was such that the scan was performed along the diameter of the germanium crystal as shown in figure\,\ref{Fig:detscan}.

\begin{figure}[t] \centering
     \includegraphics[scale=0.44]{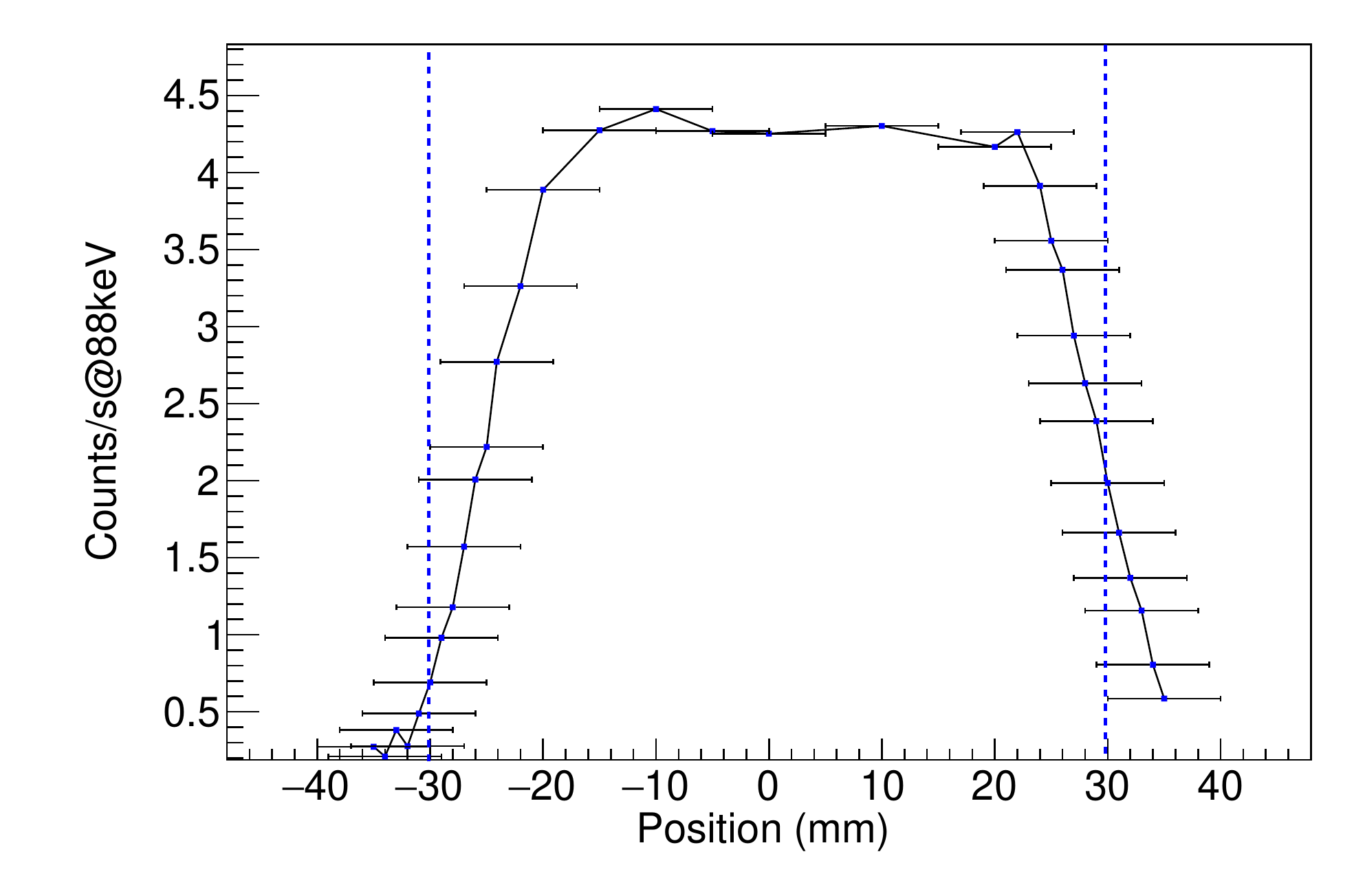}
     \caption{ \label{Fig:Hscan} Active volume scan along the cylinder diameter using an 88 keV $\gamma$-ray from the $^{109}$Cd source. The size of the error bars in the X-axis are due to the collimator aperture. The vertical blue-dashed lines indicates the crystal size of the IF-BEGe detector.}
 \end{figure}

A lead collimator with an aperture of 4
\,mm diameter and 1\,cm thickness was employed to obtain an homogeneous and focused $\gamma$-ray beam. A set of 33 measurements were taken along the horizontal axis, from left to right in steps of 1\,mm, see figure\,\ref{Fig:Hscan}. Runs of approximately one hour were taken for each position. A clear asymmetry can be seen in figure\,\ref{Fig:Hscan} between the left hand side and right hand side of the crystal with respect to its center.
This was a clear indication that the sensitive volume of the germanium crystal is indeed most likely smaller than the one indicated in the technical sheet of the detector.

In order to find a better match between the experimental data and the simulation, the following procedure was performed. First, a $^{60}$Co $\gamma$-ray source, placed 25\,cm away from the cryostat window was simulated. Most of the $\gamma$-rays from such source reaching the germanium crystal volume will pass through most of this volume \,\cite{nistGe}, if the crystal active volume is reduced in the Monte Carlo simulation, it can match the experimental data, regardless of the dead layer thickness in the front crystal face. A sensitive volume reduction of $\sim$15\,\% was found to produce good agreement between the experimental and simulated data. The second step was to find the correct position of the sensitive volume in the crystal, that yields the dead layer thickness in the front side that matches the efficiency for a source with lower energy $\gamma$-rays. Simulating a $^{133}$Ba source, also placed 25\,cm away from the cryostat window, an agreement with the experimental data fixes a dead layer of 0.05\,mm on the front face, 2.8\,mm on the back and 1.45\,mm on the sides of the crystal.

\subsection{Simulated efficiency and comparison to data}

For both detectors, the sources $^{22}$Na, $^{54}$Mn, $^{57}$Co, $^{60}$Co, $^{65}$Zn, $^{109}$Cd and $^{133}$Ba were simulated for comparison with experimental data, placed 25 cm away from the cryostat window of the detector. In the case of the IF-BEGe, the sources were simulated as ions with the GEANT4 General Particle Source and their decays with the Radioactive Decay Module\,\cite{RDMgeant4}. For the ICN-HPGe, the simulations were performed emitting mono-energetic $\gamma$-rays corresponding to the emission energy of each isotope, see table\,\ref{Tab:sources}. The deposited energy in the sensitive volume was computed adding up the deposited energy, via ionization processes, by all the secondary particles produced by the primary $\gamma$-ray. All information is stored in ROOT histograms\,\cite{BRUN199781}. For each primary $\gamma$-ray simulated, a count is stored in the corresponding deposited-energy bin. 

\begin{figure}\centering
     \includegraphics[scale=0.50]{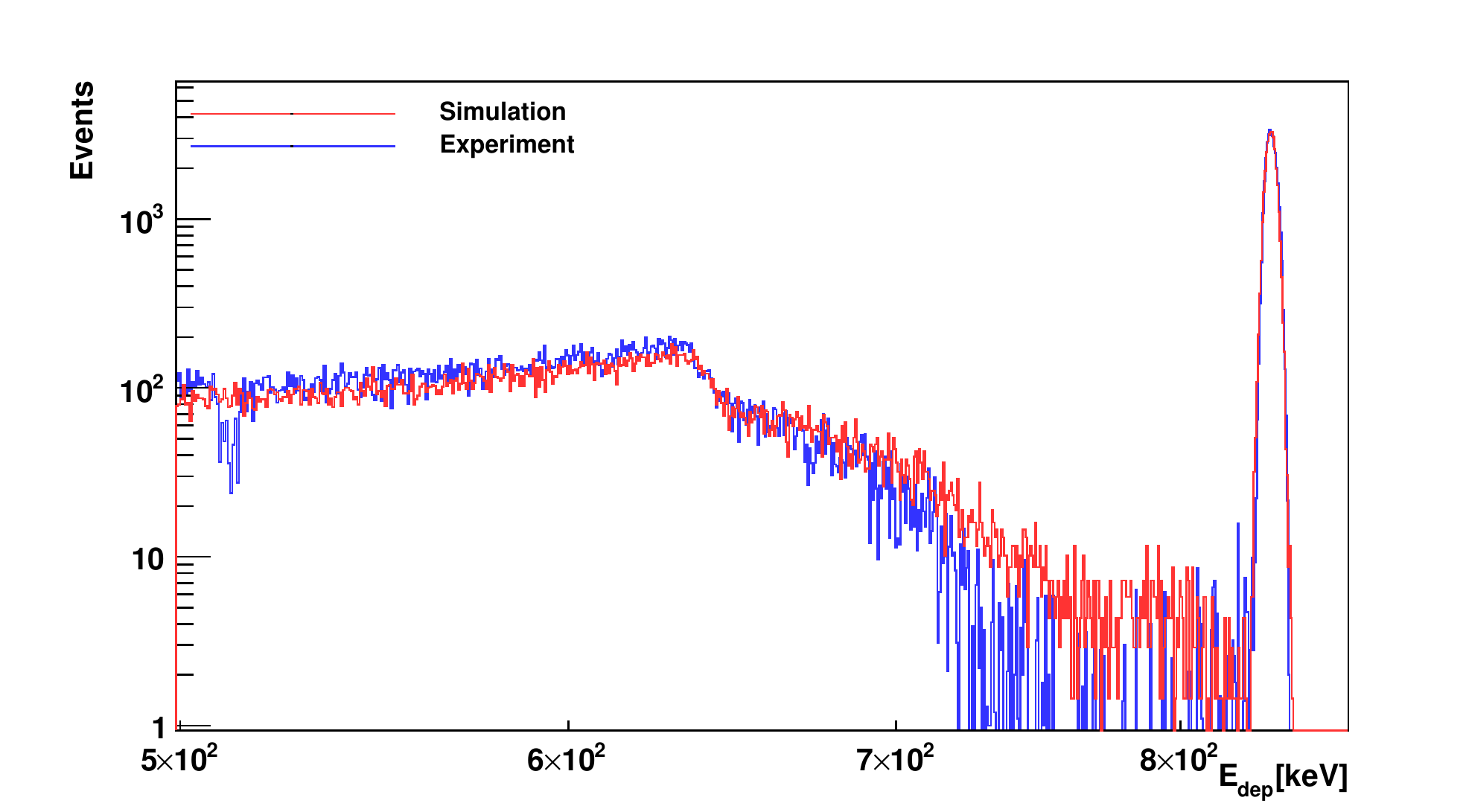}
     \caption{\label{Fig:CompSpec} Background subtracted data and Monte Carlo simulation spectra comparison for a pointlike source with $\gamma$-rays at 835\,keV (corresponding to a $^{54}$Mn emission) pointlike source, located 25\,cm away from the ICN-HPGe detector frontal face. An arbitrary normalization is used for qualitative comparison purposes.}
 \end{figure}

Figure\,\ref{Fig:CompSpec} shows a comparison between a calibrated spectrum of a $^{54}$Mn source and the Monte Carlo simulation for a pointlike source emitting $\gamma$-rays at 853\,keV. The simulated spectrum reproduces with good agreement the experimental energy spectrum features such as energy resolution, photopeak, Compton continuum and Compton edge:  maximum energy which can be transferred to an electron\,\cite{GGilmoure}, escape peak at 511\,keV has been removed by background subtraction. The energy resolution was introduced in the simulation following the empirical function obtained from experimental data, presented in section\,\ref{sec:res}.

The full energy photopeak efficiencies are computed as in section\,\ref{sec:efficiency}, equation\,\ref{eq:eff}, comparing the number of counts in the fitted Gaussian photopeak to the original number of simulated primary $\gamma$-rays. The comparison to the experimental data is shown in figure\,\ref{Fig:HPGeEff}. For both detectors, the efficiencies from the simulations are in agreement with the experimental efficiencies within uncertainties. The shapes of the curves drawn by the experimental data and Monte Carlo simulations of the detectors reflect the differences between these two. The curve for the ICN-HPGe is what is expected from a p-type coaxial detector, having a maximum in efficiency around 100\,keV. On the other hand, the IF-BEGe detector does not show a maximum, as expected from a planar type detector \,\cite{GGilmoure}. The efficiency of both detectors decreases with the energy since the higher energy $\gamma$-rays need more sensitive volume for multiple interactions to be fully absorbed.

\section{\label{sec:ext} Validation with extended calibration sources.}

In order to measure the activity in a given sample, the efficiency calibration with pointlike $\gamma$-ray sources is not sufficient, since full energy peak efficiency is a geometrically dependent quantity. Instead, an extended source efficiency is required, for the geometry of a given extended source, e.g. a sample vial.  This is described for each detector in the following subsections. For practical reasons, each detector was validated with different extended sources; for the ICN-HPGe detector, a potassium chloride (KCl) solution was used while for the IF-BEGe detector, a $^{210}$Pb solution was used. 

\subsection{ICN-HPGe detector characterization validation}

The simulation of the ICN-HPGe detector was validated using water samples with a salt substitute containing KCl at different concentrations. 
\begin{figure}[h] \centering
     \includegraphics[scale=0.25]{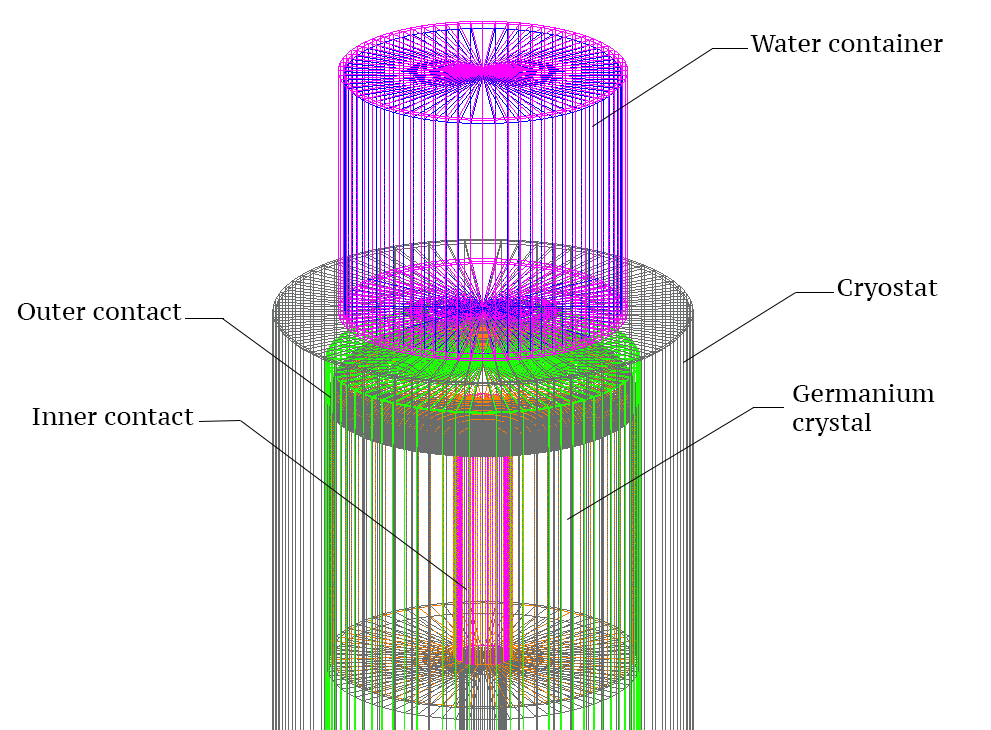}
     \includegraphics[scale=0.44]{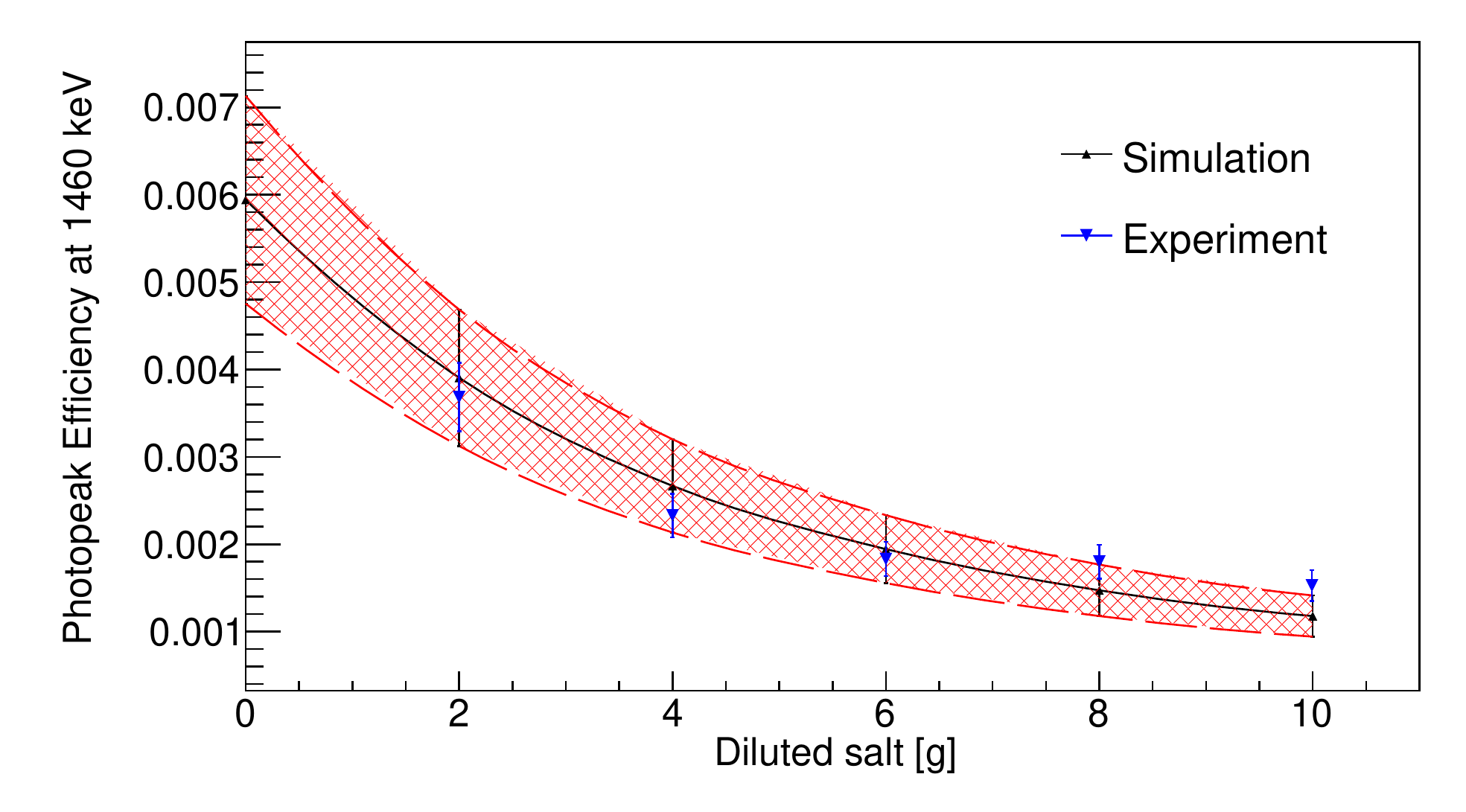}
     \caption{Top panel, geometry of the simulation of the ICN-HPGe detector with an extended source (container in pink) on top. Bottom panel, Simulated and experimental data extended source, $^{40}$K photopeak efficiency in the ICN-HPGe detector in a potassium salt solution, as a function of the  diluted salt mass. A 20\,\% systematic uncertainty is estimated in the simulation.}\label{Fig:ExtSourceICN}
 \end{figure}

Five samples of Novoxal brand KCl based salt substitute diluted in injectable water (sterile water used for medical applications) at different concentrations were prepared in cylindrical polypropylene containers of 600\,ml volume. Figure\,\ref{Fig:ExtSourceICN} (top) shows the sample container geometry together with the ICN-HPGe cryostat in the Monte Carlo simulation.

In order to determine the $^{40}$K isotope concentration per gram of salt, a background spectrum of the container with no diluted salt is recorded and subtracted from the spectra of the salt solution samples. Then, the remaining number of counts in the photopeak is determined following the same methodology as in the pointlike source case.

The geometry of the sample was implemented in the simulation, as shown in figure~\ref{Fig:ExtSourceICN} (top). 
The $\gamma$-rays emission from active $^{40}$K nuclei within the sample was
   incorporated in the simulation by sampling random points inside the
   physical volume of the solution and shooting mono-energetic and
   isotropic $\gamma$-rays with energy 1460\,keV, the most probable $\gamma$-rays channel
   of $^{40}$K. The efficiency is calculated as the ratio of the
   number of $\gamma$-rays counted inside the $^{40}$K photopeak to those initially
   emitted. As the concentration approaches zero, the number of emitter
   centers from the solution approaches zero, but the efficiency approaches a finite value, corresponding to the 
probability of detecting 1460\,keV $\gamma$-rays emitted from within the 
container inner volume when it is filled only with water. 

%The $\gamma$-ray emission from active $^{40}$K nuclei within the sample was incorporated in the simulation by sampling a point inside the physical volume of the solution and sending a mono-energetic isotropic $\gamma$-ray. Its energy is fixed to 1460\,keV, the most probable $\gamma$-ray channel of $^{40}$K.

Higher concentrations of KCl salt lead to a self-absorption effect of the $\gamma$-rays in the samples.
This effect is clearly seen in the computed 1460 keV photopeak efficiencies for different concentrations, as shown in figure\,\ref{Fig:ExtSourceICN} (bottom). With these simulated efficiencies the activity of the extended sources were determined to obtain the masses of $^{40}$K in the samples.

From the experimental data and Monte Carlo simulations, the concentration ratio of $^{40}$K per mass of diluted potassium salt (PS) was obtained as $\eta_{s} = 0.0324 \pm 0.0028\hspace{0.1cm} mg[^{40}K]/g$[PS]. From the Novoxal label information, in which there is 309.33 mg of pure potassium per salt gram, and taking into account the natural abundance of the $^{40}$K (0.012\,\%), the ratio $\eta_{l} = (0.0361 \pm 0.0018) \hspace{0.1cm} mg[^{40}K]/g$[PS] was obtained. Both results are compatible within errors, validating the measurement and the Monte Carlo simulation. This comparison assumes an uncertainty of 5\,\% of the potassium concentration in the salt and a 20\,\% systematic uncertainty is estimated in the simulated  efficiency.

\subsection{IF-BEGe detector $^{210}$Pb calibrated samples}

For the Monte Carlo simulation validation of the IF-BEGe detector, five calibrated liquid samples of $^{210}$Pb in a solution of deionized water were used. This samples were prepared at Royal Holloway, University of London (RHUL) and measured with a broad energy germanium detector in Boulby Underground Germanium Suite (BUGS)\,\cite{Scovell:2017srl} in the UK. These samples were shipped to Mexico and independently measured with the IF-BEGe detector.

The response of the Boulby BEGe P-type detector, with a 60\% relative efficiency and 0.9\,kg crystal weight used in this study is characterized using the method discussed in\,\cite{Scovell:2017srl}. An extended IAEA-385\,\cite{IAEA-385} sample with known $^{210}$Pb contamination is placed on the front face of the detector and the response to this is used to tune a GEANT4 simulation of the experimental setup. Using the simulation, the detector efficiency was determined at the 46.5 keV $^{210}$Pb peak. The extended sources were assayed on the Boulby BEGe P-type detector and their activities were determined. 

The activities measured at BUGS are compared with those measured with the IF-BEGe detector, as a cross-check to evaluate the IF-BEGe detector performance and also to validate the Monte Carlo simulation of the extended source.

\begin{figure}[h] \centering
     \includegraphics[scale=0.24]{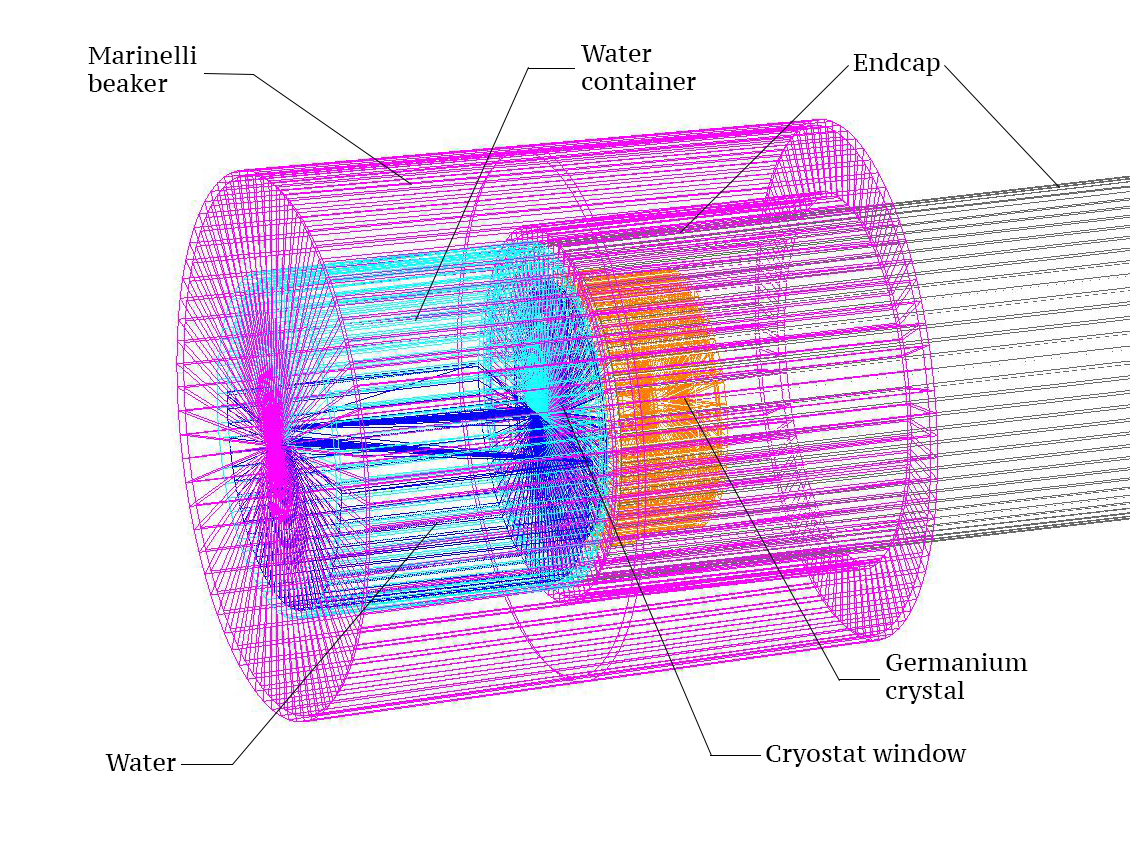}
     \caption{GEANT4 simulated geometry of the $^{210}$Pb calibrated sample inside of a polyethylene Marinelli beaker, placed in the front face of the IF-BEGe detector.} \label{Fig:Pbwater}
 \end{figure}

The IF-BEGe efficiency at lower energies was evaluated using X-rays from a pointlike $^{109}$Cd source. The agreement between the experimental data and Monte Carlo simulations is 10\,\% at 25\,keV.

For the measurement of the $^{210}$Pb calibrated sample using the IF-BEGe detector, the containers were held inside of a Marinelli beaker, which maximizes the solid-angle coverage of the detector by the sample (both containers made out of polyethylene). The experimental setup was enclosed by a 14\,cm thick lead shield, purged with the nitrogen gas from the detector dewar. After $\sim$\,24 hours of exposure time, the 46.5\,keV peak was clearly identified. The $\gamma$-rays were counted as in section\,\ref{sec:efficiency}. 

Figure \,\ref{Fig:Pbwater} shows the Monte Carlo simulation of the experimental configuration of the IF-BEGe detector with the $^{210}$Pb calibrated sample. Since the position of the water container inside the Marinelli beaker could be shifted during experiment, as the lid of the beaker was closed, a systematic uncertainties study was performed. The nominal value for the measurement was taken in the position in which the axes of symmetry of the beaker and water container are aligned among themselves and with the detector symmetry axis. Then, the position of the water container in the simulation was shifted up $+4$\,mm and down $-4$\,mm with respect to the Marinelli beaker (which was the maximum space available for movement between the containers), maintaining their axes parallel. From this study, a 14\,\% systematic uncertainty was obtained.

\begin{table}[h!]\centering
\begin{tabular}{c|c|c|c|c}
\hline
  & Mass & Expected & Boulby & IF-BEGe\\ 
     & (g) & (Bq/kg) & (Bq/kg)(stat) & (Bq/kg)(stat+sys) \\
 \hline \hline
I & 170.8 &blank & N/A & N/A  \\ \hline
II &165.6 & 40& 26.9(6) & 26(4) \\ \hline
III &177.9&22  & 23.3(3) & 17(2) \\ \hline
 IV & 163.6& N/A  & 0.029(7) & 2.5(4) \\ \hline 
 V &124.6 &  70 & 77(1) & 71(10)  \\ \hline
VI &120.45 &744 & 754(3) & 837(124) \\ \hline
\end{tabular}
\caption{Lead-210 (46.5 keV) calibration standard samples measured with IF-BEGe detector. The third column lists the specific activities (Bq/kg) expected from the solution preparation at RHUL, in the fourth column those measured at BUGS\,\cite{Scovell:2017srl}, and the last column shows the measurement reported using the IF-BEGe detector.}
\label{Tab:210Pb}
\end{table}

With the efficiency computed from the Monte Carlo simulation and the $\gamma$-ray events from the experimental data, the activities of each $^{210}$Pb sample were computed. The results are listed in table\,\ref{Tab:210Pb}, showing a good agreement with the activity values measured at BUGS, in the UK. {Row I in table\,\ref{Tab:210Pb} shows a control water sample where no radioactive material was incorporated, in this sample there was no identifiable signal at 46.5\,keV when counting in the IF-BEGe detector. }
 
\section{\label{sec:conc}Conclusions}

The High Purity Germanium detectors described in this work are part of the detector suite planned for the first underground laboratory in Mexico. The energy linearity, energy resolution and full energy peak efficiency of the detectors, in conjunction with the low radioactive background provided by the mine environment, will allow the study of a wide range of samples with low radioactive content.
These detectors will be used for experiments in nuclear and astroparticle physics, biology and geology. Other applications of the facility in mining heritage, among other areas of the physical and social sciences are under development.

The energy scale and resolution of the ICN-HPGe and IF-BEGe detectors were obtained over the energy range relevant to low-background astroparticle experiments and environmental radiation studies. The detection efficiency was quantified in this energy range using a suite of radioactive sources. From a scan to the IF-BEGe crystal, an alteration in its active volume was found. Simulations considering this alteration were performed, to find the optimal configuration, matching the measured detector efficiencies within uncertainties. The detector characterization and simulation results were validated through tests with externally-calibrated KCl and $^{210}$Pb sources. This work presents the first step towards establishing the performance of the LABChico radio-assay capabilities.

\acknowledgments
This work is supported by the STFC Global Challenges Research Fund (Foundation Awards, Grant ST/R002908/1), DGAPA UNAM grants PAPIIT-IT100420 and PAPIIT-IN108020, CONACyT grants CB-240666 and A1-S-8960. The authors thank Hesiquio Vargas Hern\'andez, technician at the machine shop at IFUNAM. They also thank Juan Estrada, Kevin Kuk and Andrew Sonnenschein for their support in the Fermilab donation of germanium detectors and instrumentation.

%\begin{thebibliography}{*99}
\bibliographystyle{unsrt}
\bibliography{HPGe_paper1}
%\end{thebibliography}

\end{document}